\documentclass{aastex63}

\received{}
\revised{}
\accepted{}
\submitjournal{ApJ}

\shorttitle{Chemistry in IRAS 16562--3959}
\shortauthors{Taniguchi et al.}

\begin{document}

\title{Chemical composition in the IRAS 16562--3959 high-mass star-forming region}

\correspondingauthor{Kotomi Taniguchi}
\email{kotomi.taniguchi@gakushuin.ac.jp}

\author[0000-0003-4402-6475]{Kotomi Taniguchi}
\affiliation{Department of Physics, Faculty of Science, Gakushuin University, Mejiro, Toshima, Tokyo 171-8588, Japan}

\author{Andr{\'e}s ~E. Guzm{\'a}n}
\affiliation{National Astronomical Observatory of Japan (NAOJ), National Institutes of Natural Sciences, Osawa, Mitaka, Tokyo 181-8588, Japan}

\author{Liton Majumdar}
\affiliation{School of Earth and Planetary Sciences, National Institute of Science Education and Research, HBNI, Jatni 752050, Odisha, India}

\author[0000-0003-0769-8627]{Masao Saito}
\affiliation{National Astronomical Observatory of Japan (NAOJ), National Institutes of Natural Sciences, Osawa, Mitaka, Tokyo 181-8588, Japan}
\affiliation{Department of Astronomical Science, School of Physical Science, SOKENDAI (The Graduate University for Advanced Studies), Osawa, Mitaka, Tokyo 181-8588, Japan}

\author[0000-0002-2062-1600]{Kazuki Tokuda}
\affiliation{Department of Physical Science, Graduate School of Science, Osaka Prefecture University, 1-1 Gakuen-cho, Naka-ku, Sakai, Osaka 599-8531, Japan}
\affiliation{National Astronomical Observatory of Japan (NAOJ), National Institutes of Natural Sciences, Osawa, Mitaka, Tokyo 181-8588, Japan}



\begin{abstract}
We have analyzed the Atacama Large Millimeter/submillimeter Array (ALMA) cycle 2 data of band 6 toward the G345.4938+01.4677 massive young protostellar object  (G345.5+1.47 MYSO) in the IRAS 16562--3959 high-mass star-forming region with an angular resolution of $\sim 0\farcs3$, corresponding to $\sim 760$ au.
We spatially resolve the central region which consists of three prominent molecular emission cores.
A hypercompact (HC) \ion{H}{2} region (Core A) and two molecule-rich cores (Core B and Core C) are identified using the moment zero images of the H30$\alpha$ line and a CH$_{3}$OH line, respectively.
Various oxygen-bearing complex organic molecules (COMs), such as (CH$_{3}$)$_{2}$CO and CH$_{3}$OCHO, have been detected toward the positions of Core B and Core C, while nitrogen-bearing species, CH$_{3}$CN, HC$_{3}$N and its $^{13}$C isotopologues, have been detected toward all of the cores. 
We discuss the formation mechanisms of H$_{2}$CO by comparing the spatial distribution of C$^{18}$O with that of H$_{2}$CO.
The $^{33}$SO emission, on the other hand, shows a ring-like structure surrounding Core A, and it peaks on the outer edge of the H30$\alpha$ emission region.
These results imply that SO is enhanced in a shock produced by the expanding motion of the ionized region.
\end{abstract}



\section{Introduction} \label{sec:intro}

Complex organic molecules (COMs), consisting of more than six atoms and rich in hydrogen, are abundant in the compact ($\sim 0.1$ pc) dense and hot gas ($n \geq 10^{7}$ cm$^{-3}$, $T > 100$ K) around massive young protostars \citep{2009ARA&A..47..427H}. 
Their chemistry is known as hot-core chemistry.
Development of radio observational facilities has enabled detailed investigation of this chemistry toward high-mass star-forming regions \citep{2019A&A...628A..27B,2019A&A...631A.142G,2019A&A...624L...5P,2018A&A...620L...6T}.
Well-studied Galactic hot cores are the Sagittarius B2 \citep[e.g.,][]{2017A&A...604A..60B,2019A&A...628A..27B} and the Orion region \citep[e.g.,][]{2012ApJS..201...17F, 2012ApJS..201...16W, 2015A&A...581A..71F}.
Outside the Galaxy, oxygen-bearing COMs (e.g., methanol, methyl formate, and dimethyl ether) have been detected from hot cores in the Large and Small Magellanic Clouds (LMC and SMC) \citep{2019ECS.....3.2088S}.

The detection in Sgr B2 of iso-propyl cyanide, a branched alkyl molecule, suggested a possible link between the interstellar COMs and pre-biotic molecules such as amino acids detected in meteorites due to their key side chain structure \citep{2014Sci...345.1584B}.
The levels of deuterium fractionation of COMs in Sgr B2 are lower than a prediction by a chemical model \citep{2016A&A...587A..91B}.
Nitrogen-bearing COMs and oxygen-bearing COMs show different spatial distributions in the Orion region \citep{2015A&A...581A..71F}. 
The chemical diversity around massive young stellar objects (MYSOs) were also found; carbon-chain-poor/COMs-rich sources and a carbon-chain-rich/COMs-poor source \citep{2017ApJ...844...68T, 2018ApJ...866...32T, 2018ApJ...866..150T}.
These results show that hot core chemistry processes are far from being settled, and detailed observational data from a variety of  sources is still needed.

The G345.5+1.47 MYSO \citep{2007AA...476.1019M} is located in the center of the IRAS 16562--3959 high-mass star-forming region \citep{2006AJ....131.1163S}.
The distance and bolometric luminosity ($L_{\rm {bol}}$) are 2.4 kpc and 154400 $L_{\sun}$ \citep{2013ApJS..208...11L}, respectively.
In the IRAS 16562--3959 high-mass star-forming region, 18 continuum cores have been detected in ALMA band 3 \citep{2014ApJ...796..117G}.
They also detected various sulfur-bearing species such as SO, SO$_{2}$, CS, and OCS, and velocity gradients in the first moment maps of the first two.
A hot molecular outflow driven by an ionized jet has been detected at the G345.5+1.47 MYSO \citep{2011ApJ...736..150G}.
\citet{2017AA...602A..59C} investigated whether circumstellar rotating disk around the G345.5+1.47 MYSO exists by analyzing the CH$_{3}$CN ($12-11$), $^{13}$CH$_{3}$CN ($13-12$), and SiO ($5-4$) rotational transitions using ALMA. 
Though such disks are ubiquitous in low-mass YSOs, they could not identify it around this high-mass YSO.
\citet{2018ApJS..236...45G} showed the spatial distributions of several molecules toward the IRAS 16562--3959 by using the ALMA band 3 observations.
In this region, \citet{2018ApJS..236...45G} suggested the presence of different chemically evolutionary stages, and thus it is crucial to investigate its chemical properties in detail.
However, an angular resolution of 1\farcs7 was not enough to spatially resolve the central G345.5+1.47 MYSO \citep{2018ApJS..236...45G}.
It is still unclear whether chemical properties of hot cores in the same region are similar to each other or not, and, if they are different, whether the difference relates to the local physical conditions or not.

In this paper, we present ALMA band 6 data toward the G345.5+1.47 MYSO with a higher spatial resolution of $\sim$ 0\farcs3 ($\sim 760$ au).
In Section \ref{sec:data}, details about the archival data and reduction procedures are described.
The continuum image, moment zero images of several molecular emission lines, and spectra toward three cores which we identify in this paper are presented in Sections \ref{sec:momzero}, \ref{sec:momzero2}, and \ref{sec:spectra}, respectively.
Analytical methods and results of the spectra are described in Section \ref{sec:ana}.
We compare the chemical composition between the two main cores detected in Section \ref{sec:dischem}. 
We discuss relationships of spatial distributions between C$^{18}$O and H$_{2}$CO, and between $^{33}$SO and H30$\alpha$ in Sections \ref{sec:dis1} and \ref{sec:dis2}, respectively.
In Section \ref{sec:con}, we summarize the main conclusions of this paper.

\section{Data and Reduction Procedure} \label{sec:data}

We present archival data from cycle 2 data, band 6 (project ID; 2013.1.00489.S, PI; Riccardo Cesaroni).
Observational details are given by \citet{2017AA...602A..59C}.
Table \ref{tab:spw} summarizes the frequency band and resolution of each spectral window.
The spectral window with the widest frequency coverage was used for a continuum observation.
The field of view (FoV) and largest angular scale (LAS) of the 12 m array observations are $\sim$ 26\arcsec and $\sim$ 4\farcs4, respectively.
The coordinate of the center of the target source is ($\alpha_{\rm{J}2000}$, $\delta_{\rm{J}2000}$) = ($16^{\rm {h}}59^{\rm {m}}$41\fs61, -40\degr03\arcmin43\farcs3).
The angular resolutions are approximately 0\farcs32$\times$0\farcs25, corresponding to 768 au $\times$ 600 au at the source distance of 2.4 kpc.

\begin{deluxetable}{cccc}
\tablenum{1}
\tablecaption{Summary of spectral windows covered by the correlator setup \label{tab:spw}}
\tablewidth{0pt}
\tablehead{
\colhead{Frequency\tablenotemark{a}} & \colhead{Frequency\tablenotemark{a}} & \colhead{Velocity} & \colhead{$1\sigma$ noise}\\
\colhead{(GHz)} & \colhead{resolution} & \colhead{resolution} & \colhead{(mJy/beam)} \\
\colhead{} & \colhead{(kHz)} & \colhead{(km s$^{-1}$)} & \colhead{} 
}
\startdata
216.976--218.849 & 1953.1 & 3.0 & 1.1 \\
219.533--219.767 & 488.3 & 0.8 & 1.5 \\
220.533--220.767 & 244.1 & 0.4 & 2.0 \\
231.803--232.037 & 488.3 & 0.8 & 2.1 \\
\enddata
\tablenotetext{a}{Obtained from \citet{2017AA...602A..59C}.}
\end{deluxetable}

We conducted data reduction and imaging using the Common Astronomy Software Application \citep[CASA v 4.3.1;][]{2007ASPC..376..127M} on the pipeline-calibrated visibilities.
The data cubes were imaged with the TCLEAN task within CASA.
Natural weighting was applied.
Velocity resolutions for each spectral window and the noise levels attained are given in Table \ref{tab:spw}.
Continuum images were obtained from the data cubes using the IMCONTSUB task.
The rms noise level of the continuum image is 1.6 mJy beam$^{-1}$.

\section{Results and Analyses} \label{sec:res}

\subsection{Continuum Image and the H$_{2}$ Column Density of Identified Cores} \label{sec:momzero}

\begin{figure}[!th]
\figurenum{1}
 \begin{center}
  \includegraphics[bb=12 10 289 253, scale=1.0]{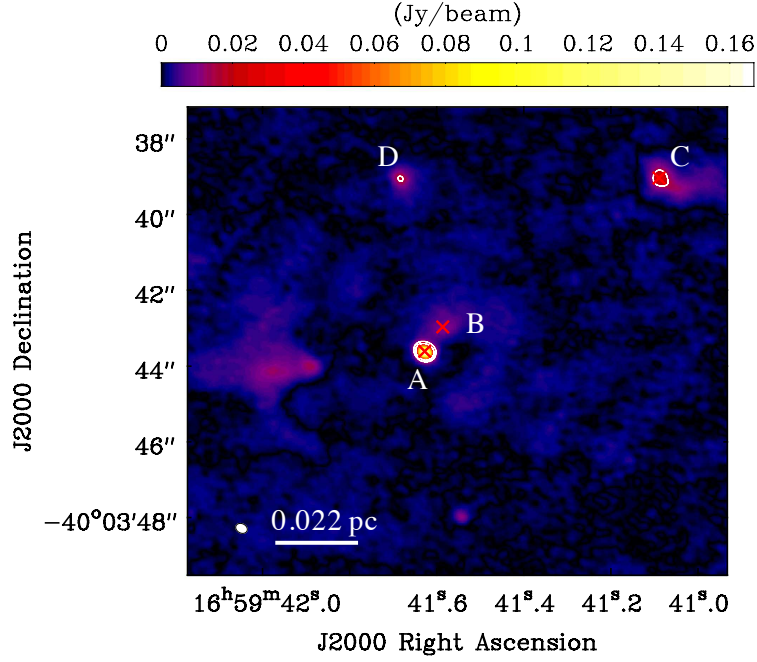} 
 \end{center}
\caption{Continuum image made from the 216.961 -- 218.834 GHz spectral window. The contour levels are 10\% and 20\% of the peak intensity (0.167 Jy beam$^{-1}$). Red crosses labeled as A -- C indicate the positions of cores identified by the H30$\alpha$ and CH$_{3}$OH moment zero images (Figure \ref{fig:core}). The white ellipse at the left bottom indicates the angular resolution of approximately 0\farcs32$\times$0\farcs25. \label{fig:continuum}}
\end{figure}

Figure \ref{fig:continuum} shows the continuum image made from the widest spectral window (216.961 -- 218.834 GHz, Table \ref{tab:spw}).
The angular resolution is 0\farcs32$\times$0\farcs25.
The strongest continuum peak associates with the G345.5+1.47 MYSO, which is indicated as Core A in Figure \ref{fig:continuum}.
Another continuum peak corresponds to Core C, which we will identify later in this subsection.
A small continuum peak, labeled as D in Figure \ref{fig:continuum}, is detected at the north position from Core A, and a weak continuum emission has been detected at the Core B position.

\begin{figure}[!th]
\figurenum{2}
 \begin{center}
  \includegraphics[bb = 20 20 267 197, scale=1.3]{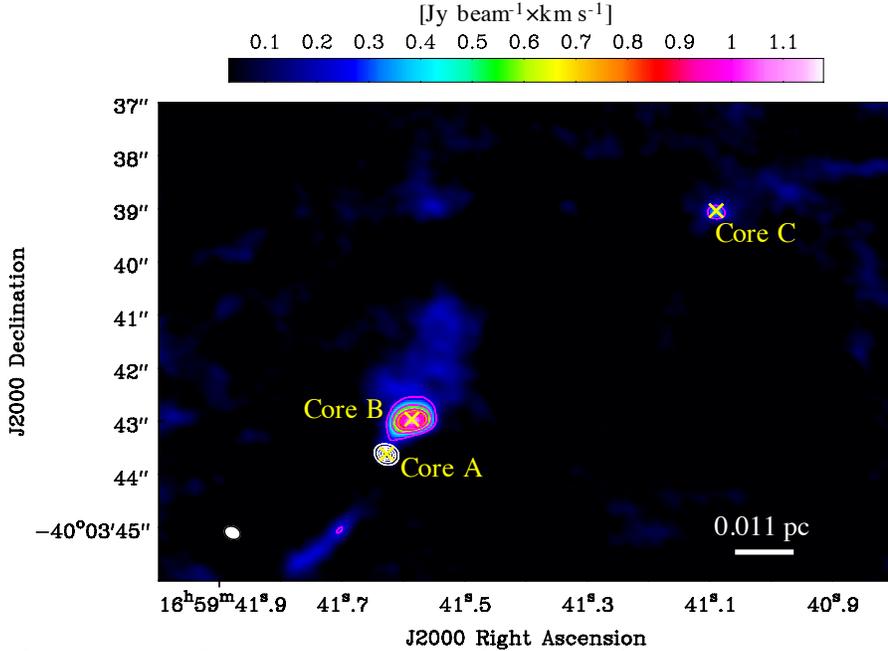} 
 \end{center}
\caption{Methanol ($4_{-2, 3}-3_{-1, 2}$ $E$) moment zero image (color image and magenta contours) overlaid by the H30$\alpha$ moment zero image (white contours). The contour levels are 20, 40, 60, and 80\% of their peak intensities, which are 1.11 and 11.1 Jy beam$^{-1} \times$ km s$^{-1}$ for the CH$_{3}$OH and H30$\alpha$ lines, respectively. Yellow crosses indicate the positions of Core A, Core B, and Core C. The white ellipse at the left bottom indicates the angular resolution of approximately 0\farcs32$\times$0\farcs25. \label{fig:core}}
\end{figure}

We define the position of the hypercompact (HC) \ion{H}{2} region and that of prominent molecular emission based on the H30$\alpha$ and CH$_{3}$OH ($4_{-2, 3}-3_{-1, 2}$ $E$) moment zero maps.
These positions are shown in the continuum map (Figure \ref{fig:continuum}) and in the moment zero maps of the H30$\alpha$ and of the methanol transition shown in Figure \ref{fig:core}.
Core A corresponds to the HC\ion{H}{2} region, and Cores B and C correspond to the strongest rich molecular cores identified in IRAS 16562--3959 by \citet{2017AA...602A..59C} and by \citet{2018ApJS..236...45G}.
Table \ref{tab:coreid} lists general properties of these cores based on 2D Gaussian fittings to the moment zero maps.

We derived the H$_{2}$ column density, $N_{\rm{H}_{2}}$, at each core using the following formula \citep{2008A&A...487..993K}:
\begin{equation} \label{equ:H2}
N_{\rm{H}_{2}} = 2.02 \times 10^{20} \Bigl(e^{1.439(\lambda /{\rm {mm}})^{-1}(T/10 {\rm {K}})^{-1}}-1\Bigr)\Bigl(\frac{\kappa_{\nu}}{0.01\: {\rm {cm}}^{2}\: {\rm {g}}^{-1}}\Bigr)^{-1}\Bigl(\frac{F_{\nu}^{\rm {beam}}}{{\rm {mJy}}\: {\rm {beam}}^{-1}}\Bigr)\Bigl(\frac{\theta_{\rm {HPBW}}}{10\: {\rm {arcsec}}}\Bigr)^{-2}\Bigl(\frac{\lambda}{{\rm {mm}}}\Bigr)^{3},
\end{equation}
where 
\begin{equation} \label{equ:kappa}
\kappa_{\nu} = 0.1 \Bigr(\frac{\nu}{1 {\rm {THz}}}\Bigl)^{\beta}
\end{equation}
The symbols of $\lambda$, $T$, $\kappa_{\nu}$, $F_{\nu}^{\rm {beam}}$, and $\theta_{\rm {HPBW}}$ in Equation (\ref{equ:H2}) are the wavelength, temperature, dust opacity, flux of dust continuum emission, and half power beam width, respectively.
The continuum data wavelength is approximately 1.37 mm.
The flux values at each core are summarized in Table \ref{tab:coreid}.
These fluxes are beam-averaged values with beam sizes of 0\farcs3 for Core A and 0\farcs5 for the others, respectively.
We corrected the continuum emission toward Core A by subtracting the free-free continuum emission as we describe in Appendix \ref{sec:Halpha}.
We subtracted this free-free component from the continuum flux and derived the dust continuum emission.
H30$\alpha$ emission toward Core B and Core C was not detected.
The dust opacity at the 1.37 mm ($\kappa_{\nu}$) is calculated using Equation (\ref{equ:kappa}) \citep{2011A&A...536A..23P} with $\beta=1.6$ \citep{2013ApJ...767..126S}.
The dust temperature ($T$) is assumed to be 100 K, because COMs have been detected toward all of the cores as shown in Section \ref{sec:momzero2}.
If we change the temperature between 50 and 200 K, the derived $N_{\rm{H}_{2}}$ values change by a factor of $\sim2$.
The derived $N_{\rm{H}_{2}}$ values are summarized in Table \ref{tab:coreid}.

\begin{deluxetable*}{cccccccc}
\tablenum{2}
\tablecaption{Identification of cores\label{tab:coreid}}
\tablewidth{0pt}
\tablehead{
\colhead{Position} & \colhead{R.A. (J2000)} & \colhead{Decl. (J2000)} & \colhead{$\theta_{\rm {major}} \times \theta_{\rm {minor}}$} & \colhead{Position Angle} & \colhead{$V_{\rm {sys}}$ (km s$^{-1}$)\tablenotemark{a}} & \colhead{Flux (mJy beam$^{-1}$)\tablenotemark{b}} & \colhead{$N_{\rm{H}_{2}}$ (cm$^{-2}$)\tablenotemark{c}}
}
\startdata
Core A & $16^{\rm {h}}59^{\rm {m}}$41\fs627 &  -40\degr03\arcmin43\farcs61 & 0\farcs30 $\times$ 0\farcs25 & 66\degr & -17.1 &  $119.0 \pm 1.6$ & $5.0^{+5.5}_{-2.6} \times 10^{24}$ \\
Core B & $16^{\rm {h}}59^{\rm {m}}$41\fs586 &  -40\degr03\arcmin42\farcs96 & 0\farcs70 $\times$ 0\farcs48 & 108\degr & -14.6 & $7.8 \pm 1.6$ & $1.9^{+2.1}_{-1.0} \times 10^{23}$\\
Core C & $16^{\rm {h}}59^{\rm {m}}$41\fs089 &  -40\degr03\arcmin39\farcs03 & 0\farcs54 $\times$ 0\farcs47 & 27\degr & -11.8 &  $19.2 \pm 1.6$ & $4.8^{+5.3}_{-2.4} \times 10^{23}$ \\
\enddata
\tablecomments{Core A is identified based on the moment zero map of the H30$\alpha$ line by the 2D Gaussian fitting. Core B and Core C are identified by the 2D Gaussian fitting of the moment zero image of the CH$_{3}$OH ($4_{-2, 3}-3_{-1, 2}$ $E$) line.}
\tablenotetext{a}{Obtained by the Gaussian fitting of the $J=12-11$, $K=4$ line of CH$_{3}$CN.}
\tablenotetext{b}{Continuum fluxes with beam sizes of 0\farcs3 for Core A and 0\farcs5 for the others, respectively.}
\tablenotetext{c}{Errors were derived by changes in assumed temperatures between 50 and 200 K.}
\end{deluxetable*}

\clearpage
\subsection{Moment Zero Images} \label{sec:momzero2}

\begin{figure}[!th]
\figurenum{3}
 \begin{center}
  \includegraphics[bb = 0 10 410 400, scale=1.0]{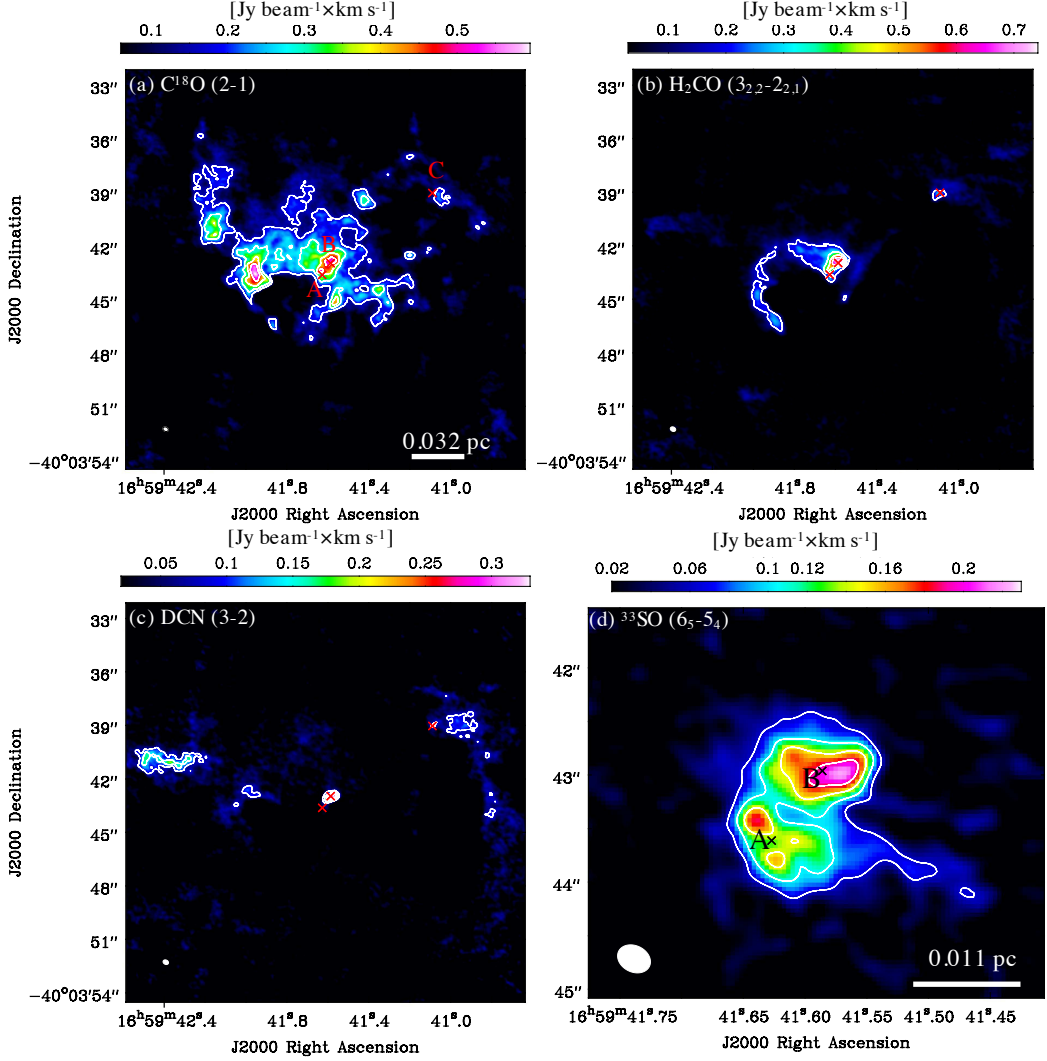} 
 \end{center}
\caption{Moment zero images of (a) C$^{18}$O ($2-1$), (b) H$_{2}$CO ($3_{2, 2}- 2_{2, 1}$), (c) DCN ($3-2$), and (d) $^{33}$SO ($6_{5}-5_{4}$). The contour levels are 20, 40, 60, and 80\% of their peak intensities (0.60, 0.74, 0.33, and 0.28 Jy beam$^{-1}$$\times$ km s$^{-1}$ for panels (a) -- (d), respectively).
The white ellipse at the left bottom indicates the angular resolution of approximately 0\farcs32$\times$0\farcs25. 
Red crosses in panels (a) -- (c) indicate positions of Cores A, B, and C. 
In panel (d), black crosses show positions of Core A and Core B.
The color scales are adjusted from rms noise levels to the peak intensities for each panel.
\label{fig:f1}}
\end{figure}

\begin{figure}[!th]
\figurenum{4}
 \begin{center}
  \includegraphics[bb= 0 15 499 528, scale=1.0]{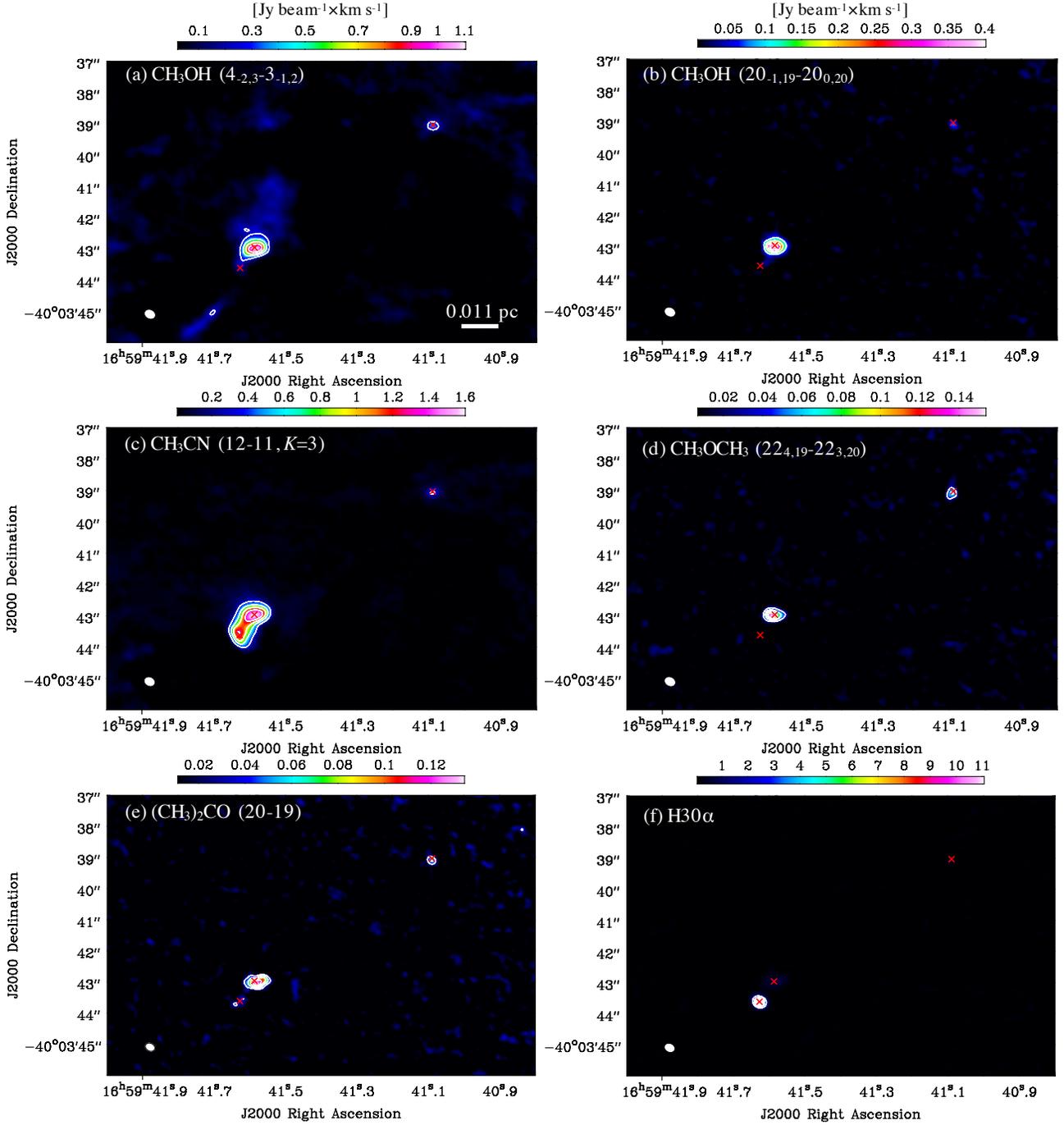} 
 \end{center}
\caption{Moment zero images. The contour levels are 20, 40, 60, and 80\% of their peak intensities. The full transitions and peak intensities of each panel are summarized in Table \ref{tab:map}. The white ellipse at the left bottom indicates the angular resolution of approximately 0\farcs32$\times$0\farcs25. Red crosses indicate positions of Cores A, B, and C. The color scales are adjusted from rms noise levels to the peak intensities for each panel. \label{fig:f2}}
\end{figure}

\begin{figure}[!th]
\figurenum{5}
 \begin{center}
  \includegraphics[bb= 0 22 525 658, scale=1.0]{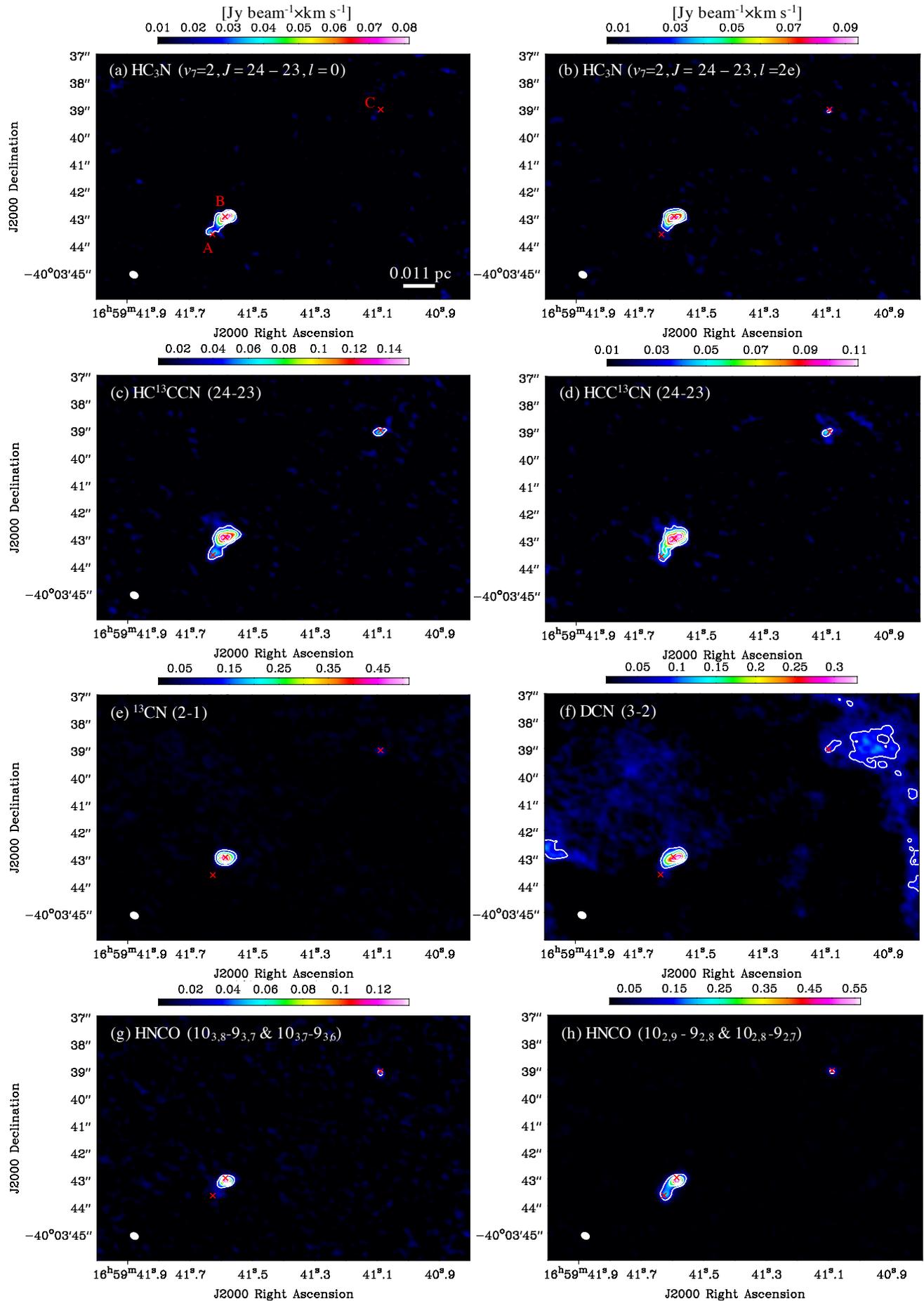} 
 \end{center}
\caption{Moment zero images. The contour levels are 20, 40, 60, and 80\% of their peak intensities. The transitions and peak intensities of each panel are summarized in Table \ref{tab:map}. The white ellipse at the left bottom indicates the angular resolution of approximately 0\farcs32$\times$0\farcs25. Red crosses indicate positions of Cores A, B, and C. The color scales are adjusted from rms noise levels to the peak intensities for each panel. \label{fig:f3}}
\end{figure}

We made moment zero images of each molecular line integrating the whole velocity range where emission is detected.
Figure \ref{fig:f1} shows the spatial distributions of (a) C$^{18}$O ($2-1$), (b) H$_{2}$CO ($3_{2, 2}- 2_{2, 1}$), (c) DCN ($3-2$), and (d) $^{33}$SO ($6_{5}-5_{4}$).
Their spatial distributions, except for $^{33}$SO, are relatively extended compared to those of COMs (Figures \ref{fig:f2} and \ref{fig:f3}).
The red crosses indicate positions of Cores A to C (Figure \ref{fig:core} and Table \ref{tab:coreid}).
Table \ref{tab:map} summarizes species, transition, rest frequency, excitation energy, and peak intensity of each moment zero image analyzed in this work.

The spatial distributions of C$^{18}$O and H$_{2}$CO show arc-like structures with strong peaks at Core B.
There is also weaker emission associated with Core C.
In Section \ref{sec:dis1}, we discuss the formation mechanisms of H$_{2}$CO by comparing between the spatial distributions of C$^{18}$O and H$_{2}$CO.

The DCN spatial distribution (panel (c) of Figure \ref{fig:f1}) is different from the above two species.
A strong peak is located at Core B and a filamentary structure can be seen at the northern east region, corresponding to the northeast outflow cavity wall (NEC-wall) in \citet{2018ApJS..236...45G}.
Other emission regions are located nearby and to the west of Core C.

Panel (d) of Figure \ref{fig:f1} shows a close-up moment zero image of $^{33}$SO ($6_{5}-5_{4}$) emission at G345.5+1.47.
One $^{33}$SO peak is associated with Core B, and other peaks are located near Core A.
The $^{33}$SO emission seems to surround Core A.
We discuss the $^{33}$SO spatial distribution around Core A in detail in Section \ref{sec:dis2}.

\begin{deluxetable}{cclccc}
\tablenum{3}
\tablecaption{Summary of moment zero images \label{tab:map}}
\tablewidth{0pt}
\tablehead{
\colhead{Panel} & \colhead{Species} & \colhead{Transition} & \colhead{Rest Frequency} & \colhead{$E_{\rm {u}}/k$} & \colhead{Peak intensity} \\
\colhead{} & \colhead{} & \colhead{} & \colhead{(GHz)} & \colhead{(K)} & \colhead{(Jy beam$^{-1} \times$ km s$^{-1}$)}
}
\startdata
Figure \ref{fig:f1} & & & & & \\
(a) & C$^{18}$O & $J=2-1$ & 219.5603541 & 15.8 & 0.60 \\
(b) & H$_{2}$CO & $J_{Ka, Kc}=3_{2, 2}- 2_{2, 1}$ &218.475632 & 68.1 & 0.74 \\
(c) & DCN & $J=3-2$ & 217.2385378 & 20.9 & 0.33 \\
(d)\tablenotemark{a} & $^{33}$SO & $J_{N}=6_{5}-5_{4}$, $F=\frac{9}{2}-\frac{7}{2}$ & 217.8271782 & 34.7 & 0.28 \\
     &                  & $J_{N}=6_{5}-5_{4}$, $F=\frac{11}{2}-\frac{9}{2}$ & 217.8298337 & 34.7 & ... \\
     &                  & $J_{N}=6_{5}-5_{4}$, $F=\frac{13}{2}-\frac{11}{2}$ & 217.8317691 & 34.7 & ... \\    
     &                  & $J_{N}=6_{5}-5_{4}$, $F=\frac{15}{2}-\frac{13}{2}$ & 217.8326422 & 34.7 & ... \\  
Figure \ref{fig:f2} & & & & & \\    
(a) & CH$_{3}$OH & $4_{-2, 3}-3_{-1, 2}$ $E$ & 218.440063 & 45.5 & 1.11 \\
(b) & CH$_{3}$OH & $20_{-1,19}-20_{-0,20}$ $E$ & 217.886504 & 508.4 & 0.40 \\
(c)\tablenotemark{b} & CH$_{3}$CN & $J=12-11$, $K=3$ & 220.7090165 & 133.2 & 1.60 \\
(d)\tablenotemark{a} & CH$_{3}$OCH$_{3}$ & $22_{4,19}-22_{3,20}$ $EA$ & 217.189668 & 253.4 & 0.15 \\
                                &                                & $22_{4,19}-22_{3,20}$ $AE$ & 217.189669 & 253.4 & ... \\
                                &				     & $22_{4,19}-22_{3,20}$ $EE$ & 217.191400 & 253.4 & ... \\
                                &				     & $22_{4,19}-22_{3,20}$ $AA$ & 217.193132 & 253.4 & ... \\ 
(e)\tablenotemark{a, b} & (CH$_{3}$)$_{2}$CO & $20_{2,18}-19_{3,17}$ $EE$ & 218.1272074 & 119.1 & 0.13 \\
				    & 				    & $20_{3,18}-19_{3,17}$ $EE$ & 218.1272074 & 119.1 & ... \\
				    &					    & $20_{2,18}-19_{2,17}$ $EE$ & 218.1272074 & 119.1 & ... \\ 
				    &					    & $20_{3,18}-19_{2,17}$ $EE$ & 218.1272074 & 119.1 & ... \\ 	
(f) & H30$\alpha$ & ... & 231.900928 & ... & 11.1 \\		
Figure \ref{fig:f3} & & & & & \\ 	
(a) & HC$_{3}$N & $v_{7}=2$, $J=24-23$, $l=0$ & 219.6751141 & 773.5 & 0.082 \\
(b) & HC$_{3}$N & $v_{7}=2$, $J=24-23$, $l=2e$ & 219.7073487 & 776.8 & 0.095 \\        
(c) & HC$^{13}$CCN & $v=0$, $J=24-23$ & 217.3985682 & 130.4 & 0.15 \\
(d) & HCC$^{13}$CN & $v=0$, $J=24-23$ & 217.4195740 & 130.4 & 0.11 \\          
(e)\tablenotemark{a} & $^{13}$CN & $N= 2-1$, $J=\frac{3}{2}-\frac{1}{2}$, $F_{1}=2-1$, $F=1-0$ & 217.296605 & 15.6 & 0.51\\
                                &                  & $N= 2-1$, $J=\frac{5}{2}-\frac{3}{2}$, $F_{1}=2-2$, $F=2-2$ & 217.298937 & 15.7 & ...\\
                                &		    & $N= 2-1$, $J=\frac{3}{2}-\frac{1}{2}$, $F_{1}=2-1$, $F=2-1$ & 217.301175 & 15.6 & ...\\
                                &		    & $N= 2-1$, $J=\frac{3}{2}-\frac{1}{2}$, $F_{1}=2-1$, $F=3-2$ & 217.303191 & 15.6 & ...\\           
(f) & DCN & $J=3-2$ & 217.2385378 & 20.9 & 0.33 \\          
(g)\tablenotemark{a} & HNCO & $10_{3,8}-9_{3,7}$ & 219.6567695 & 433.0 & 0.14\\
				    &		    & $10_{3,7}-9_{3,6}$ & 219.6567708 & 433.0 & ...\\
(h)\tablenotemark{a} & HNCO & $10_{2,9}-9_{2,8}$ & 219.7338500 & 228.3 & 0.56\\
				    & 	    & $10_{2,8}-9_{2,7}$ & 219.7371930 & 228.3 & ... \\				                                            
\enddata
\tablecomments{Rest frequency and excitation energy are taken from the Cologne Database for Molecular Spectroscopy \citep[CDMS;][]{2005JMoSt.742..215M}.}
\tablenotetext{a}{These lines were not resolved and detected as one line.}
\tablenotetext{b}{Rest frequency and excitation energy are taken from the Jet Propulsion Laboratory catalog \citep[JPL catalog;][]{1998JQSRT..60..883P}.}
\end{deluxetable}

Figures \ref{fig:f2} and \ref{fig:f3} show moment zero maps of various molecular emission lines more compact morphology.
The information of lines and peak intensities for each panel is listed in Table \ref{tab:map}.
Most of the molecular emissions are associated with both Cores B and C as shown in Figures \ref{fig:f2} and \ref{fig:f3}.

Figure \ref{fig:f2} shows the spatial distributions of COMs in panels (a) -- (e) and H30$\alpha$ in panel (f).
The CH$_{3}$OH ($4_{-2, 3}-3_{-1, 2}$ $E$) line in panel (a) shows more extended spatial distribution than the CH$_{3}$OH ($20_{-1,19}-20_{-0,20}$ $E$) line in panel (b).
This is caused by their different upper energy levels; $E_{\rm {u}}/k = 44.5$ K for panel (a) and 508.4 K for panel (b), respectively (Table \ref{tab:map}).

The CH$_{3}$CN ($12-11$, $K=3$) line comes from Cores A, B, and C, while the lines of oxygen-bearing COMs mainly come from Core B and Core C (Figure \ref{fig:f2}). 
The upper energy level of the CH$_{3}$CN line of panel (c) is 133.2 K, which is an intermediate value compared to those of panels (a) and (b).
Therefore, the different spatial distributions between CH$_{3}$OH and CH$_{3}$CN are not brought by the different upper energy level of the lines.
The different morphologies between the CH$_{3}$OH ($4_{-2, 3}-3_{-1, 2}$ $E$) and the CH$_{3}$CN line seem to reflect different origins.
As indicated in Figure \ref{fig:line1}, the CH$_{3}$OH ($4_{-2, 3}-3_{-1, 2}$ $E$) line has been detected at Core A, although there is no its emission peak at Core A (panel (a) of Figure \ref{fig:f2}). 
Since low excitation energy lines of CH$_{3}$OH trace shock regions \citep[e.g.,][]{2020MNRAS.493.2395T}, the CH$_{3}$OH line around Core A is possibly originated in a shock region.
Shock regions are induced by several star formation phenomena, such as jets and molecular outflows. 
\citet{2011ApJ...736..150G} identified the SE-NW molecular outflow and 
the central knot of the jet/outflow system is consistent with the Core A position \citep{2010ApJ...725..734G}.
Alternatively, the CH$_{3}$OH line may trace a remnant gas scattered by the HC\ion{H}{2} region.
The CH$_{3}$CN emission, on the other hand, does peak at Core A, which implies thermal sublimation from dust grains.

The spatial distribution of the CH$_{3}$OCH$_{3}$ ($22_{4,19}-22_{3,20}$) lines in panel (d) is similar to that of (CH$_{3}$)$_{2}$CO ($20-19$) lines in panel (e).
The upper energy levels of the lines in panels (d) and (e) are 253.4 K and 119.1 K, respectively (Table \ref{tab:map}).
Both of these lines likely trace hot core regions.
In addition, their distributions at Core B are more compact than that of the CH$_{3}$OH ($4_{-2, 3}-3_{-1, 2}$ $E$; $E_{\rm {u}}/k = 44.5$ K) line in panel (a), and similar to the CH$_{3}$OH ($20_{-1,19}-20_{-0,20}$ $E$; $E_{\rm {u}}/k = 508.4$ K) line in panel (b).
These different spatial distributions seem to reflect lower abundances of CH$_{3}$OCH$_{3}$ and (CH$_{3}$)$_{2}$CO compared to the CH$_{3}$OH abundance (Section \ref{sec:ana}) and different upper energy levels.

Figure \ref{fig:f3} shows the spatial distributions of HC$_{3}$N and its $^{13}$C isotopologues ($24-23$) in panels (a) -- (d), $^{13}$CN ($2-1$) in panel (e), DCN ($3-2$) in panel (f), and HNCO ($10-9$) in panels (g) and (h).
The vibrationally excited lines of HC$_{3}$N ($v_{7}=2$, $J=24-23$) mainly come from Core B, while the ground vibrational state lines ($v=0$, $J=24-23$) of the $^{13}$C isotopologues associate with Cores A, B, and C, as shown in panels (a) -- (d).
The two $^{13}$C isotopologues of HC$_{3}$N, HC$^{13}$CCN and HCC$^{13}$CN, show the same spatial distributions in panels (c) and (d).
The more extended distributions of the $^{13}$C isotopologues are caused by the lower energy levels of the observed lines.
In fact, these vibrationally excited lines have extremely high upper energy levels of $\sim 775$ K, compared to those of the ground vibrational state lines (130.4 K) in panels (c) and (d).
The HC$_{3}$N spatial distributions are similar to that of CH$_{3}$CN ($12-11$), a typical hot core tracer.

The $^{13}$CN ($2-1$) emission in panel (e) is associated with Core B and its weak emission comes from Core C, while the DCN ($3-2$) distribution in panel (f) shows a more extended structure.
The upper energy levels are 15.6 K and 20.9 K for the $^{13}$CN and DCN lines, respectively.
Because these are quite similar to each other, different upper energy transition levels is not the main cause of the different spatial distributions of these two species.
This difference is more likely to be caused by the different environments traced by each species.
The CN/HCN ratio is enhanced in high ultraviolet (UV) flux regions, because HCN is destroyed by the UV radiation forming CN \citep[e.g.,][]{2018MNRAS.481.4662R}.
A possible explanation for the non-detection of the $^{13}$CN line at Core A, at which the UV flux is likely strongest in this region, is the selective photodissociation.
In order to investigate the effect of the selective photodissociation, we need to observe the H$^{13}$CN lines and lines of normal species of CN and HCN.

Four HNCO ($10-9$) lines also show similar spatial distributions to CH$_{3}$CN and HC$_{3}$N; the HNCO lines associate with Cores A, B, and C as shown in panels (g) and (h).
The differences in spatial distribution between panels (g) and (h) arises from the different upper energy levels.
The upper energy levels of lines in panel (g) are 433.0 K and higher than those in panel (h) (228.3 K), and therefore, the spatial distribution of panel (h) is more extended.

The critical density is another important factor to determine spatial distributions of each line.
Since we could not derive accurate densities at each position, we do not discuss the critical density.
However, we note that most molecular lines have been detected at Core B, at which the derived H$_{2}$ column density is lower than the other cores.
This suggests the upper energy level of the transition is more relevant than the critical density to explain the emission and make the compositions presented in this study.

\clearpage
\subsection{Spectra toward the three cores} \label{sec:spectra}

\begin{figure}[!th]
\figurenum{6}
 \begin{center}
  \includegraphics[bb = 0 30 402 639, scale=0.85]{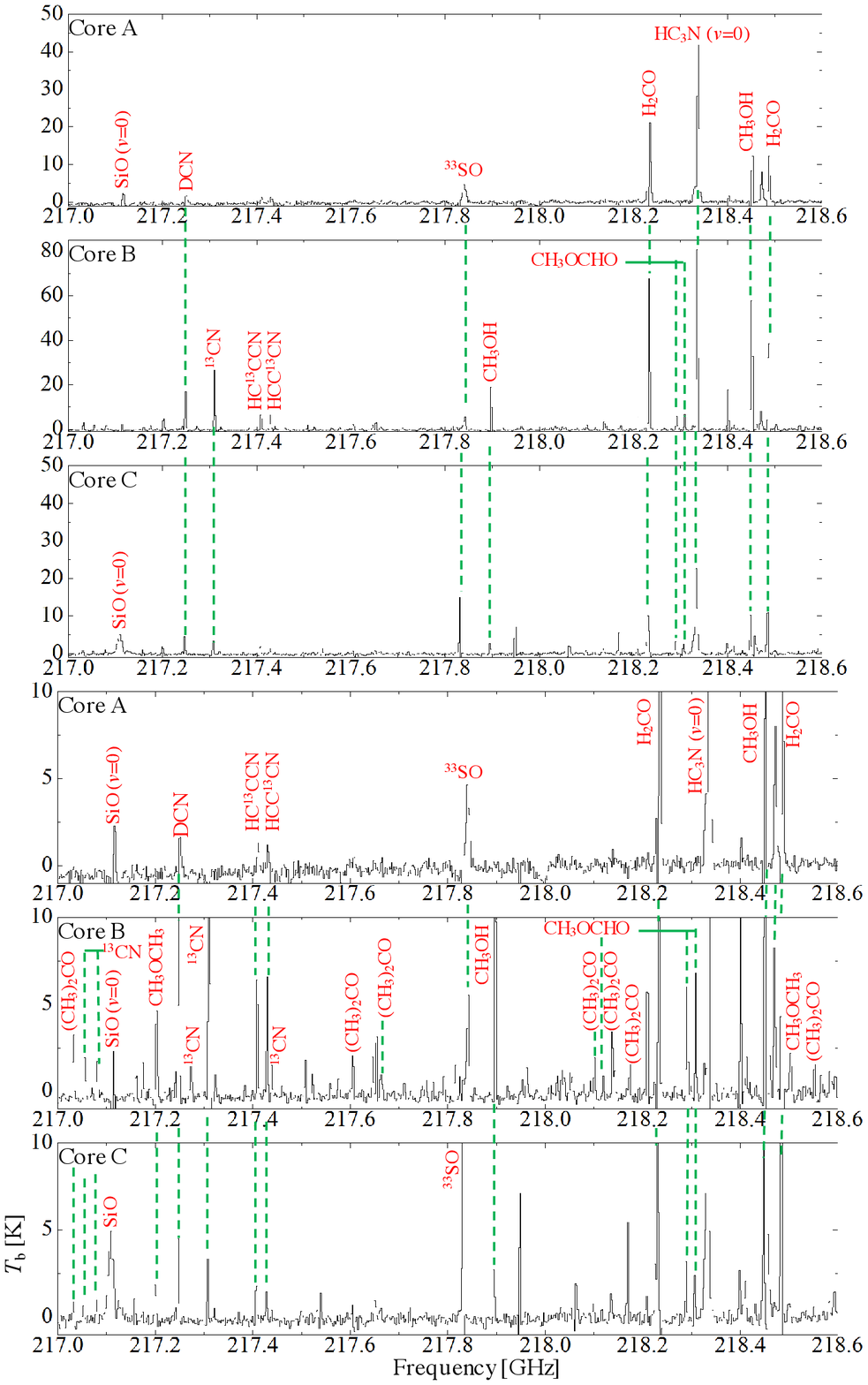} 
 \end{center}
\caption{Spectra in the frequency range of 217.0--218.4 GHz toward the three cores. Bottom three panels are a zoom of the top three.\label{fig:line1}}
\end{figure}

\begin{figure}[!th]
\figurenum{7}
 \begin{center}
  \includegraphics[bb = 0 30 431 577, scale=0.8]{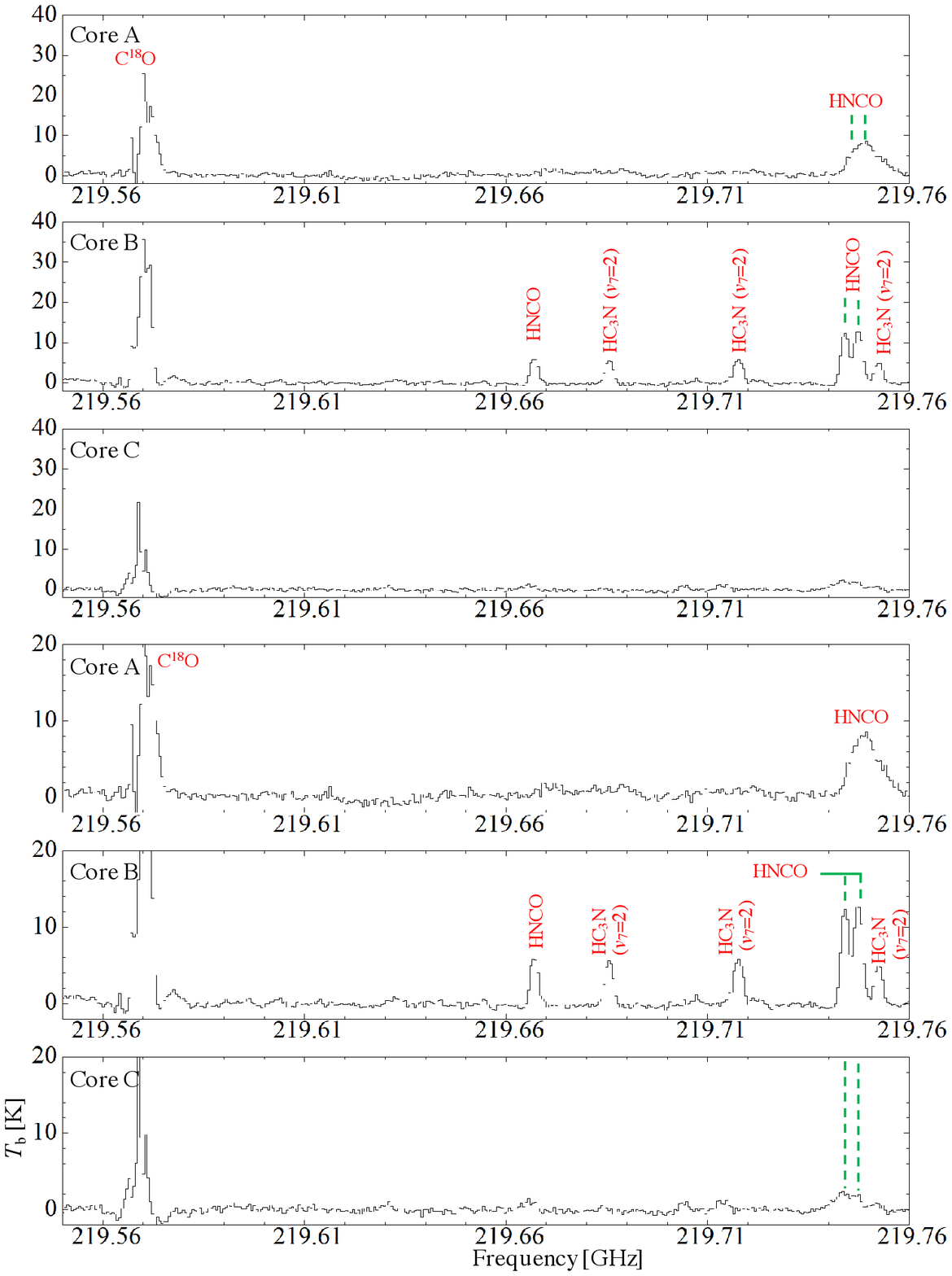} 
 \end{center}
\caption{Spectra in the frequency range of 219.55--219.76 GHz toward the three cores. Bottom three panels are a zoom of the top three.\label{fig:line2}}
\end{figure}

Figures \ref{fig:line1} and \ref{fig:line2} show spectra of the 217.0 -- 218.6 GHz and 219.56 -- 219.76 GHz bands toward Cores A, B, and C, respectively.   
Bottom three panels are a zoom of the top three.
We identified lines using the CASSIS software with the Cologne Database for Molecular Spectroscopy \citep[CDMS;][]{2005JMoSt.742..215M} and the Jet Propulsion Laboratory catalog \citep[JPL catalog;][]{1998JQSRT..60..883P}.
Since the velocity resolution of the final spectra of Figure \ref{fig:line1} is low (3 km s$^{-1}$), some lines could not be identified without ambiguity.
At the position of Core C, the $^{33}$SO line is detected as a spurious-like line (Figure \ref{fig:line1}).
Such a spurious-like feature is considered to be brought by very low-velocity resolution (3 km s$^{-1}$). 
We confirm that it does not affect for other regions.
We then neglect the position of Core C in its moment zero image.
We did not apply Gaussian fitting for detected lines due to the low velocity resolution.
Table \ref{tab:line} summarizes species, transition, rest frequency, and excitation energy of detected lines at each core position.

Core B is the most line-rich position, where various oxygen-bearing COMs, e.g., CH$_{3}$OCH$_{3}$ and CH$_{3}$OCHO, have been detected.
Furthermore, vibrational-excited lines of HC$_{3}$N, whose upper energy levels are extremely high ($E_{\rm u}/k \approx 775$ K), have been detected.

As shown later, the excitation temperature of CH$_{3}$CN is derived to be above 200 K at Core C.
Furthermore, the abundances of COMs are high.
For example, the CH$_{3}$OH abundance is around $10^{-6}$ (Figure \ref{fig:chem}).
This abundance can be reproduced only after the temperature reaches above 200 K \citep{2019ApJ...881...57T}.
These results suggest active hot core chemistry at Core C.
\citet{2018ApJS..236...45G} also concluded that Core C is associated with a second hot molecular core within IRAS16562--3959, linked to an MYSO less massive than G345.5+1.47.
In summary, all of these results imply that a very young star is embedded at Core C.

\begin{deluxetable}{clccccc}
\tablenum{4}
\tablecaption{Summary of detected lines \label{tab:line}}
\tablewidth{0pt}
\tablehead{
\colhead{Species} & \colhead{Transition} & \colhead{Rest Frequency} & \colhead{$E_{\rm {u}}/k$} & \colhead{Core A\tablenotemark{a}} & \colhead{Core B\tablenotemark{a}} & \colhead{Core C\tablenotemark{a}} \\
\colhead{} & \colhead{} & \colhead{(GHz)} & \colhead{(K)} & \colhead{} & \colhead{} & \colhead{}
}
\startdata
(CH$_{3}$)$_{2}$CO & $19_{3,16}-18_{4,15}$, $19_{4,16}-18_{3,15}$ & 217.0225 & 115.5 & N & Y & (Y) \\
$^{13}$CN  & $N=2-1$, $J=3/2-3/2$, $F_{1}=1-1$, $F=1-2$ & 217.046988 & 15.7 & N & Y & (Y) \\
                  & $N=2-1$, $J=3/2-3/2$, $F_{1}=1-1$, $F=2-1$ & 217.0728010 & 15.7 & N & Y & (Y) \\
SiO & $v=0$, $J=5-4$ & 217.104980 & 31.3 & Y & (Y) & Y \\     
CH$_{3}$OCH$_{3}$\tablenotemark{b} & $22_{4,19}-22_{3,20}$ $EA$ & 217.189668 & 253.4 & N & Y & Y \\
				    & $22_{4,19}-22_{3,20}$ $AE$ & 217.189669 & 253.4 & N & Y & Y \\
                               & $22_{4,19}-22_{3,20}$ $EE$ & 217.191400 & 253.4 & N & Y & Y \\
                               & $22_{4,19}-22_{3,20}$ $AA$ & 217.193132 & 253.4 & N & Y & Y \\ 
DCN & $J=3-2$ & 217.2385378 & 20.9 & Y & Y & Y \\     
$^{13}$CN & $N=2-1$, $J=3/2-1/2$, $F_{1}=1- 0$, $F=0-1$ & 217.264639 & 15.7 & N & Y & N \\
	          &  $N=2-1$, $J=5/2-3/2$ & 217.296605 & 15.7 & N & Y & Y \\         
	          & $N=2-1$, $J=5/2-3/2$, $F_{1}=2-2$, $F=2-3$ & 217.315147 & 15.7 & N & Y & N \\      
HC$^{13}$CCN & $J=24-23$ & 217.3985682 & 130.4 & Y & Y & Y \\
HCC$^{13}$CN & $J=24-23$ &  217.419574 & 130.4 & Y & Y & Y \\	    
$^{13}$CN & $N= 2-1$, $J=5/2-3/2$, $F_{1}= 2-1$, $F=3-2$ & 217.4285632 & 15.7 & N & Y & (Y) \\ 
(CH$_{3}$)$_{2}$CO & $18_{12,6}-17_{13,4}$ & 217.5921139 & 141.3 & N & N & (Y) \\
                               & $23_{20,3}-22_{21,1}$\tablenotemark{b} & 217.6552 & 248.3 & N & Y & (Y) \\
                               & $37_{12,25}-37_{11,26}$,  $37_{13,25}-37_{12,26}$\tablenotemark{b} & 217.6553 & 493.2 & N & Y & (Y) \\
$^{33}$SO\tablenotemark{b} & $J_{N}=6_{5}-5_{4}$, $F=\frac{9}{2}-\frac{7}{2}$ & 217.8271782 & 34.7 & Y & Y & (Y) \\
                       & $J_{N}=6_{5}-5_{4}$, $F=\frac{11}{2}-\frac{9}{2}$ & 217.8298337 &  34.7 & Y & Y & (Y) \\
                       & $J_{N}=6_{5}-5_{4}$, $F=\frac{13}{2}-\frac{11}{2}$ & 217.8317691 & 34.7 & Y & Y & (Y)  \\    
                       & $J_{N}=6_{5}-5_{4}$, $F=\frac{15}{2}-\frac{13}{2}$ & 217.8326422 & 34.7 & Y & Y & (Y) \\  
CH$_{3}$OH & $20_{-1,19}-20_{-0,20}$ $E$ & 217.886504 & 508.4 & N & Y & Y \\
(CH$_{3}$)$_{2}$CO & $20_{2,18}-19_{3,17}$, $20_{3,18}-19_{2,17}$ & 218.0914 & 119.2 & N & Y & (Y) \\
				   & $32_{22,10}-32_{19,13}$ & 218.1059 & 438.9 & N & Y &  (Y) \\
				   & $20_{2,18}-19_{3,17}$, $20_{3,18}-19_{3,17}$, $20_{2,18}-19_{2,17}$ & 218.1272 & 119.1 & N & Y & (Y) \\
				   & $20_{2,18}-19_{3,17}$, $20_{3,18}-19_{2,17}$ & 218.1629 & 119.0 & N & Y & (Y) \\
H$_{2}$CO & $3_{0,3}-2_{0,2}$ & 218.2222 & 21.0 & Y & Y & Y\\
CH$_{3}$OCHO & $17_{3,14}-16_{3,13}$ & 218.2809 & 99.7 & N & Y & Y\\
			    & $17_{3,14}-16_{3,13}$ & 218.2979 & 99.7 & N & Y & Y\\ 
HC$_{3}$N & $v=0$, $J=24-23$ & 218.3247 & 131.0 & Y & Y & Y \\
CH$_{3}$OH & $4_{-2,3}-3_{-1,2}$ & 218.440063 & 45.5 & Y & Y & Y\\
H$_{2}$CO & $3_{2,2}-2_{2,1}$ & 218.4756 & 68.1 & Y & Y & Y\\
CH$_{3}$OCH$_{3}$ & $23_{3,21}-23_{2,22}$ & 218.4898 & 263.8 & N & Y & (Y)\\
 				    & $23_{3,21}-23_{2,22}$ & 218.4924 & 263.8 & N & Y & (Y)\\
				    & $23_{3,21}-23_{2,22}$ & 218.495 & 263.8 & N & Y & (Y)\\
(CH$_{3}$)$_{2}$CO & $19_{12,8}-18_{13,5}$ & 218.5439 & 154.3 & N & Y & (Y)\\				    
C$^{18}$O & $J=2-1$ & 219.5603541 & 15.8 & Y & Y & Y \\
HNCO\tablenotemark{b} & $10_{3,8}-9_{3,7}$ & 219.6567695 & 433.0 & N & Y & N\\
  	    & $10_{3,7}-9_{3,6}$ & 219.6567708 & 433.0 & N & Y & N \\
HC$_{3}$N & $v_{7}=2$, $J=24-23$, $l=0$ & 219.67465 & 769.7 & N & Y & N\\
                 & $v_{7}=2$, $J=24-23$, $l=2e$ & 219.70689 & 772.3 & N & Y & N\\
HNCO & $10_{2,9}-9_{2,8}$ & 219.7338500 & 228.3 & Y & Y & (Y)\\
         & $10_{2,8}-9_{2,7}$ & 219.7371930 & 228.3 & Y & Y & (Y) \\
HC$_{3}$N & $v_{7}=2$, $J=24-23$, $l=2f$ & 219.74174 & 772.3 & N & Y & N\\        	                                 
\enddata
\tablecomments{Rest frequency and excitation energy are taken from the Cologne Database for Molecular Spectroscopy \citep[CDMS;][]{2005JMoSt.742..215M} and the Jet Propulsion Laboratory catalog \citep[JPL catalog;][]{1998JQSRT..60..883P}.}
\tablenotetext{a}{``Y" and ``N" mean detection and non-detection at each core, respectively. The symbol of (Y) means tentative detection.}
\tablenotetext{b}{These lines were not resolved and detected as one line.}
\end{deluxetable}

\subsection{Analyses} \label{sec:ana}

\begin{figure}[!th]
\figurenum{8}
 \begin{center}
  \includegraphics[bb = 59 270 450 750, scale=0.75]{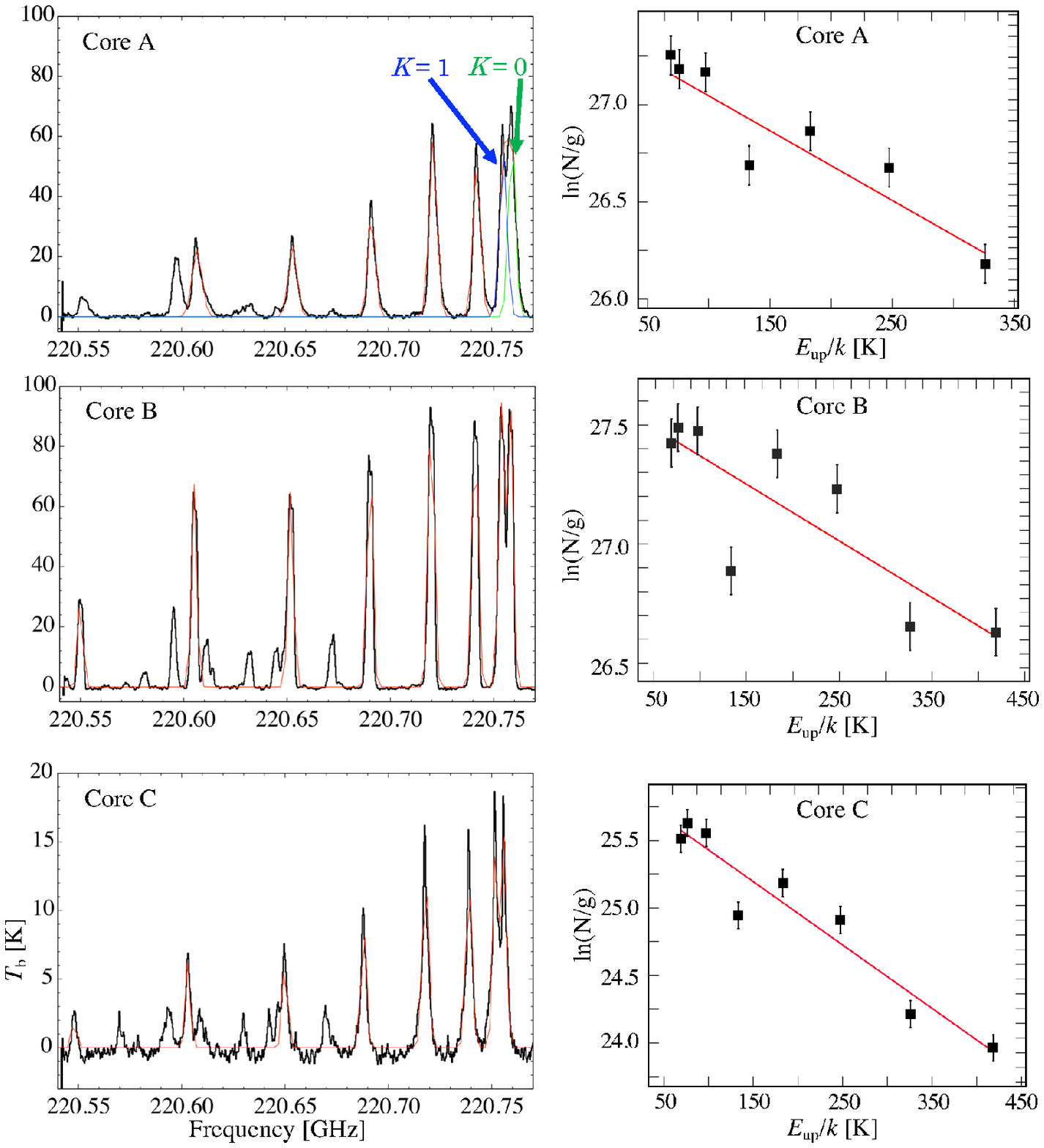}
 \end{center}
\caption{Left panels; spectra of the $J=12-11$ transition lines of CH$_{3}$CN toward Cores A, B, and C. Red lines indicate the results of the Gaussian fitting. Right panels; rotational diagrams toward Cores A, B, and C. Red lines show the fitting results, synthesized spectra of the Gaussian fitting for each line. At Core A, green and blue lines indicate the Gaussian fitting for the $K=0$ and $K=1$ lines.\label{fig:CH3CN}}
\end{figure}

We analyzed spectra at the three cores using the CASSIS software \citep{2011IAUS..280P.120C}.
In the analyses presented here, we have used the local thermodynamic equilibrium (LTE) model available in CASSIS spectrum analyzer by assuming lines are optically thin. 
The rotational diagram fittings of CH$_{3}$CN do not show any systematic shifts as shown in Figure \ref{fig:CH3CN}.
Thus, the assumption of optically thin regime seems to be reasonable in our case.

Except for CH$_{3}$CN, only one or a few lines with similar excitation energies have been detected for each species. 
We then applied the Markov Chain Monte Carlo (MCMC) method, which is an interactive process that goes through all of the parameters with a random walk and heads into the solutions space, and $\chi$$^{2}$ minimization gives the final solution.

We derive the excitation temperatures and column densities of CH$_{3}$CN using its $K-$ladder lines of the $J=12-11$ transition.
Left panels of Figure \ref{fig:CH3CN} show spectra of the eight $K-$ladder lines of the $J=12-11$ transition ($K=0-7$) toward the three cores.
The lines from right to left correspond to from $K=0$ to $K=7$.
We fitted the spectra with a Gaussian profile.
We cannot fit the $K=7$ line toward Core A, and we omitted it from the fitting.

Right panels of Figure \ref{fig:CH3CN} show the rotational diagram using the Gaussian fitting results of the left panels.
The derived column densities and excitation temperatures are summarized in Table \ref{tab:CH3CN}.
The column densities and excitation temperatures are ($1.0 \pm 0.1$)$\times 10^{16}$ cm$^{-2}$ and $279 \pm 57$ K, ($1.4 \pm 0.2$)$\times 10^{16}$ cm$^{-2}$ and $420 \pm 117$ K, and ($1.3 \pm 0.2$)$\times 10^{15}$ cm$^{-2}$ and $213 \pm 25$ K at Cores A, B, and C, respectively.
Although we assumed that the CH$_{3}$CN lines are optically thin, we cannot rule out the optically thick case due to the scatter. 
The plots for the first three $K-$ladder lines ($K=0-2$) are coincident with each other within their errors, but the plots for the $K=0$ line may be slightly lower than the other two plots. 
If the lines are optically thick, the derived rotational temperature should be overestimated \citep{2011A&A...525A.151B,2011A&A...525A..72F}.
 
\begin{deluxetable*}{ccc}
\tablenum{5}
\tablecaption{Column density and excitation temperature of CH$_{3}$CN at the three cores\label{tab:CH3CN}}
\tablewidth{0pt}
\tablehead{
\colhead{Position} & \colhead{$N$ (cm$^{-2}$)} & \colhead{$T_{\rm {ex}}$ (K)}
}
\startdata
Core A & ($1.0 \pm 0.1$)$\times 10^{16}$ & $279 \pm 57$ \\ 
Core B & ($1.4 \pm 0.2$)$\times 10^{16}$ & $420 \pm 117$ \\
Core C & ($1.3 \pm 0.2$)$\times 10^{15}$ & $213 \pm 25$ \\
\enddata
\tablecomments{Errors represent the standard deviation.}
\end{deluxetable*}

We derived the column densities and excitation temperatures of other species using the MCMC method and the LTE model in the CASSIS software.
We considered the following two cases:
\begin{enumerate}
\item assume excitation temperatures between 50 and 200 K, and
\item assume a range of temperatures around the excitation temperatures derived by the CH$_{3}$CN fitting.
\end{enumerate}
We assume the excitation temperature range of the first case taking the typical temperature at the hot core of 100 K into consideration.
For the second case, we tested the following temperature ranges: 220 -- 340 K for Core A, 250 -- 540 K for Core B and 185 -- 240 K for Core C, respectively.

We fixed source sizes, or size of the emitting region, to 0\farcs3 for Core A and 0\farcs5 for Core B and Core C, respectively (Table \ref{tab:coreid}).
Because the cores appear to be resolved, we assume a filling factor of unity.
The line width (full width half maximum; FWHM) was treated as a free parameter, constrained between 1 and 10 km s$^{-1}$.
The line widths derived by the fitting are $\sim2-4$ km s$^{-1}$ for Core A and Core B, and $\sim3-5$ km s$^{-1}$ for Core C, respectively.
Derived line widths agree well with those of typical hot cores.

Table \ref{tab:molecule} summarizes the derived column densities and excitation temperatures of the detected species except for CH$_{3}$CN toward each core.
The ``Case 1" and ``Case 2" represent the different assumed excitation temperature ranges, as described above.
The excitation temperatures of H$_{2}$CO and CH$_{3}$OH are consistent within their standard deviation errors in all of cases.
Other oxygen-bearing COMs (CH$_{3}$OCHO, CH$_{3}$OCH$_{3}$, and (CH$_{3}$)$_{2}$CO) have similar excitation temperatures, except for (CH$_{3}$)$_{2}$CO assuming Case 1 toward Core C.
Although there is a large uncertainty of the CH$_{3}$CN excitation temperature at Core B, the excitation temperatures of all the species are coincident with each other at each core in Case 2.
We will discuss comparisons of the chemical compositions between Cores B and C in Section \ref{sec:dischem} with the results of Case 2.

\begin{deluxetable*}{ccccccccc}
\tablenum{6}
\tablecaption{Derived column density and excitation temperature\label{tab:molecule}}
\tablewidth{0pt}
\tablehead{
\colhead{Species} & \multicolumn{2}{c}{Core A} & \colhead{} & \multicolumn{2}{c}{Core B} & \colhead{} & \multicolumn{2}{c}{Core C} \\
\cline{2-3} \cline{5-6} \cline{8-9} 
\colhead{} & \colhead{$N$ (cm$^{-2}$)} & \colhead{$T_{\rm {ex}}$ (K)} & \colhead{} & \colhead{$N$ (cm$^{-2}$)} & \colhead{$T_{\rm {ex}}$ (K)} & \colhead{} & \colhead{$N$ (cm$^{-2}$)} & \colhead{$T_{\rm {ex}}$ (K)} 
}
\startdata
\multicolumn{9}{c}{Case 1\tablenotemark{a}} \\
CH$_{3}$OH & ($4.3 \pm 2.3$)$\times 10^{19}$ & $98 \pm 14$ & & ($3.8 \pm 0.6$)$\times 10^{18}$ & $169 \pm 20$ & & ($4.4 \pm 4.8$)$\times 10^{18}$ & $58 \pm 4$ \\
HC$_{3}$N & ($1.6 \pm 0.7$)$\times 10^{17}$ & $135 \pm 50$ & & ($1.1 \pm 0.4$)$\times 10^{18}$ & $99 \pm 28$ & & ($1.4 \pm 0.8$)$\times 10^{18}$ & $60 \pm 9$ \\
HC$^{13}$CCN & ($3.1 \pm 0.7$)$\times 10^{14}$ & $130 \pm 37$ & & ($6.9 \pm 1.4$)$\times 10^{14}$ & $103 \pm 30$ & & ($6.1 \pm 1.9$)$\times 10^{13}$ & $76 \pm 19$ \\
HCC$^{13}$CN & ($2.4 \pm 0.6$)$\times 10^{14}$ & $130 \pm 38$ & & ($4.8 \pm 0.9$)$\times 10^{14}$ & $111 \pm 23$ & & ($1.3 \pm 0.3$)$\times 10^{14}$ & $92 \pm 32$ \\ 
H$_{2}$CO & ($1.4 \pm 0.7$)$\times 10^{17}$ & $96 \pm	37$ & & ($8.0 \pm 6.3$)$\times 10^{18}$ & $140 \pm 16$ & & ($6.2 \pm 	3.4$)$\times 10^{16}$ & $59 \pm 6$ \\
CH$_{3}$OCHO &  ... & ... & & ($8.3 \pm 1.4$)$\times 10^{16}$ & $60 \pm 6$ & & ($3.2 \pm 0.7$)$\times 10^{16}$ & $79 \pm 18$ \\
CH$_{3}$OCH$_{3}$ & ... & ... & & ($4.0 \pm 0.5$)$\times 10^{17}$ & $64 \pm 5$ & & ($2.1 \pm 0.9$)$\times 10^{17}$ & $62 \pm 11$ \\
(CH$_{3}$)$_{2}$CO & ... & ... & & ($8.3 \pm 1.0$)$\times 10^{15}$ & $66 \pm 8$ & & ($7.2 \pm 2.1$)$\times 10^{15}$ & $122 \pm 24$ \\
HNCO & ... & ... &  & ($7.0 \pm 3.6$)$\times 10^{16}$ & $82 \pm 14$ & & ($6.5 \pm 1.1$)$\times 10^{15}$ & $134 \pm 17$\\
\multicolumn{9}{c}{Case 2\tablenotemark{b}} \\
CH$_{3}$OH & ($9.6 \pm 8.9$)$\times 10^{17}$ & $256 \pm 16$ & & ($7.4 \pm 3.4$)$\times 10^{18}$ & $300 \pm 31$ & & ($3.2 \pm 1.5$)$\times 10^{17}$ & $209 \pm 11$\\ 
HC$_{3}$N & ($9.1 \pm 6.5$)$\times 10^{17}$ & $245 \pm 14$ & & ($1.2 \pm 0.9$)$\times 10^{18}$ & $278 \pm 18$ & & ($9.0 \pm 6.3$)$\times 10^{17}$ & $193 \pm 4$\\
HC$^{13}$CCN & ($3.5 \pm 0.5$)$\times 10^{14}$ & $280 \pm 35$ & & ($8.3 \pm 1.1$)$\times 10^{14}$ & $363 \pm 57$ & & ($6.8 \pm 4.7$)$\times 10^{13}$ & $192 \pm 3$\\
HCC$^{13}$CN & ($2.7 \pm 0.5$)$\times 10^{14}$ & $281 \pm 28$ & & ($6.1 \pm 1.0$)$\times 10^{14}$ & $349 \pm 59$ & & ($1.1 \pm 0.2$)$\times 10^{14}$ & $205 \pm 14$\\
H$_{2}$CO & ($1.5 \pm 0.5$)$\times 10^{17}$ & $258 \pm 30$ & & ($1.9 \pm 1.6$)$\times 10^{18}$ & $278 \pm 14$ & & ($2.2 \pm 0.4$)$\times 10^{16}$ & $204 \pm 8$\\
CH$_{3}$OCHO &  ... & ... & & ($9.5 \pm 2.2$)$\times 10^{16}$ & $277 \pm 21$ & & ($6.2 \pm 1.7$)$\times 10^{16}$ & $216 \pm 12$\\
CH$_{3}$OCH$_{3}$ & ... & ... & & ($7.9 \pm 4.0$)$\times 10^{16}$ & $272 \pm 19$ & & ($3.2 \pm 0.5$)$\times 10^{16}$ & $192 \pm 5$\\
(CH$_{3}$)$_{2}$CO & ... & ... & & ($7.8 \pm 2.6$)$\times 10^{16}$ & $264 \pm 15$ & & ($8.9 \pm 0.9$)$\times 10^{15}$ & $192 \pm 6$\\
HNCO & ... & ... &  & ($1.4 \pm 0.3$)$\times 10^{16}$ & $297 \pm 30$ & & ($4.0 \pm 0.3$)$\times 10^{15}$ & $196 \pm 6$\\
\enddata
\tablenotetext{a}{Assume that excitation temperatures are between 50 and 200 K.}
\tablenotetext{b}{Use the excitation temperatures derived by the CH$_{3}$CN fitting.}
\end{deluxetable*}

\section{Discussions} \label{sec:dis}

\subsection{Comparison of the Chemical Composition Between Core B and Core C} \label{sec:dischem}

\begin{figure}[!th]
\figurenum{9}
 \begin{center}
  \includegraphics[bb = 0 30 302 294, scale = 0.75]{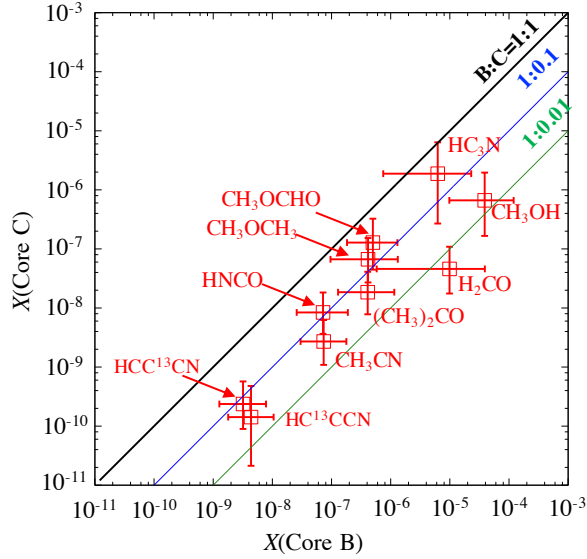}
 \end{center}
\caption{Comparison of chemical compositions between Cores B and C. The black, blue and green lines indicate $X$(Core B) = $X$(Core C), $X$(Core B) = $10 \times X$(Core C), and $X$(Core B) = $100 \times X$(Core C), respectively. \label{fig:chem}}
\end{figure}

We derived fractional abundances, $X$(molecules)=$N$(molecules)/$N_{\rm{H}_{2}}$, of each species toward Cores B and C, and compare them as shown in Figure \ref{fig:chem}.
We took the errors from both $N$(molecules) and $N_{\rm{H}_{2}}$ into consideration.

The observed H$_{2}$CO and CH$_{3}$OH abundances at Cores B and C can be reproduced after their thermal desorption from dust grains (Figure \ref{fig:H2COmodel} in Appendix \ref{sec:H2COmodel} and Figure 6 of \citet{2019ApJ...881...57T}).
The thermal desorption of H$_{2}$CO and CH$_{3}$OH occurs at $\sim 50$ K and $\sim 100$ K, respectively.
Abundances of other COMs ($\sim 10^{-8}-10^{-7}$) are reproduced in a hot core model with temperature above 100 K \citep{2013ApJ...765...60G}. 
These results suggest hot core chemistry is taking place at both Core B and Core C.

All of the species have larger fractional abundances at Core B compared to Core C.
While the fractional abundances of CH$_{3}$OH and H$_{2}$CO at Core B are higher than those at Core C by a factor of $\simeq 100$, the differences in fractional abundances of other species are around one order of magnitude.
Methanol (CH$_{3}$OH) is important for formation of oxygen-bearing COMs \citep{2009A&A...504..891O}. 
Hence, oxygen-bearing COMs except for H$_{2}$CO and CH$_{3}$OH are expected to be formed later than H$_{2}$CO and CH$_{3}$OH.
These results suggest that Core B is more chemically rich or a more evolved hot core with enough time for molecules to sublimate from dust grains and newly form in the hot gas.

\subsection{Formation Mechanisms of H$_{2}$CO} \label{sec:dis1}

Most of COMs are generally abundant in hot core regions with temperatures above 100 K where ice mantles sublimate.
As shown in Figures \ref{fig:f1} and \ref{fig:f2}, the spatial distribution of H$_{2}$CO is different from distributions of other oxygen-bearing COMs; the distribution of H$_{2}$CO shows an arc-like structure with a strong peak at Core B, while other COMs are concentrated at Core B and Core C.
Such different spatial distributions may suggest that the main formation mechanism of H$_{2}$CO is different from those of other COMs.
In this subsection, we discuss possible main formation mechanisms of H$_{2}$CO. 

According to \citet{2019ApJ...881...57T}, the formation mechanisms of H$_{2}$CO are highly sensitive to the physical evolution of the cores. 
Based on this, we can clearly distinguish three major formation processes of H$_{2}$CO (Figure \ref{fig:H2COmodel} in Appendix \ref{sec:H2COmodel}):
\begin{enumerate}
\item formation in the gas-phase through the reaction CH$_{3}$ + O $\rightarrow$ H$_{2}$CO + H, active in cold starless core phase when $T\approx10$ K and $n_{\rm{H}} \approx 10^{4}-10^{7}$ cm$^{-3}$;
\item non-thermal desorption of H$_{2}$CO ice formed through the successive hydrogenation of CO ice (ice-H + ice-CO $\rightarrow$ ice-HCO, ice-HCO + ice-H $\rightarrow$ ice-H$_{2}$CO) active in the lukewarm stage when 10 K $< T <$ 25 K and $n_{\rm{H}} \approx 10^{7}$ cm$^{-3}$;
\item thermal-evaporation of H$_{2}$CO ice, active in hot-core stage when $T > 50$ K and $n_{\rm{H}} \approx 10^{7}$ cm$^{-3}$.
\end{enumerate}

\begin{figure}[!th]
\figurenum{10}
 \begin{center}
  \includegraphics[bb = 0 18 249 190, scale=1.1]{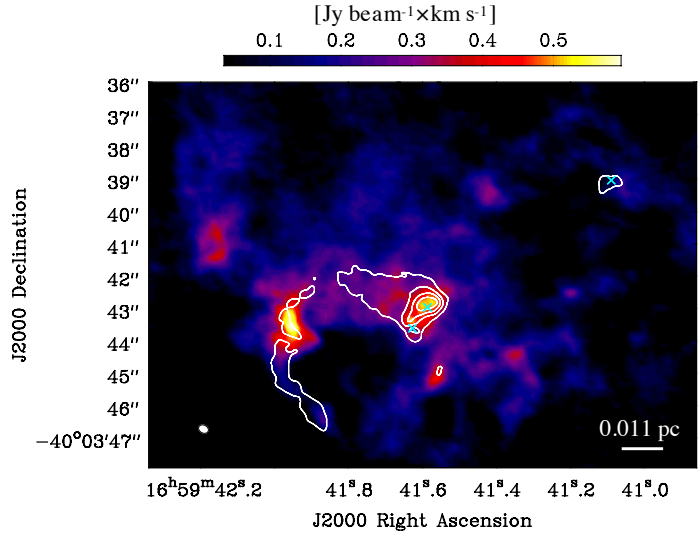} 
 \end{center}
\caption{Comparison of spatial distributions of C$^{18}$O (color) and H$_{2}$CO (white contours; the contour levels are 20, 40, 60, and 80\% of its peak intensity of 0.74 Jy beam$^{-1}$$\times$ km s$^{-1}$). Cyan crosses indicate positions of Cores A, B, and C. \label{fig:f4}}
\end{figure}

Figure \ref{fig:f4} shows the spatial distributions of C$^{18}$O (color) and H$_{2}$CO (white contours).
Their spatial distributions show a similar structure; an arc-like extended structure and a strong peak at Core B.
The extended structure seems to suggest that heating sources are not necessary for formation of the gas-phase H$_{2}$CO. 
This means that either gas-phase formation or non-thermal desorption is important for H$_{2}$CO. 
The similar spatial distributions of C$^{18}$O and H$_{2}$CO support the grain-surface formation and non-thermal desorption. 
Still, it does not mean that gas-phase formation is irrelevant for H$_{2}$CO.
To distinguish the two reactions, we need further observations, e.g., \ion{O}{1} distribution.

The H$_{2}$CO abundance with respect to H$_{2}$ at Core B is derived to be around $10^{-5}$ (Figure \ref{fig:chem}).
This abundance can be reproduced only after thermal evaporation of H$_{2}$CO (Figure \ref{fig:H2COmodel}) with temperatures above 50 K.
Since many COMs have been detected at Core B, the dust temperature is expected to be above 100 K. 
Thus, the thermal evaporation seems to work efficiently at this position.
 
\subsection{Spatial distribution of $^{33}$SO emission around Core A} \label{sec:dis2}

\citet{2014ApJ...796..117G} showed the spatial distributions of sulfur-bearing species, SO, SO$_{2}$, CS, and OCS, with an angular resolution of $\sim$ 2\farcs3 $\times$ 1\farcs3.
They found that the emission lines of these species come from a molecular core with a size of around 3000 au at the G345.5+1.47 MYSO.
\citet{2014ApJ...796..117G} suggested that the observed SO emission and morphology at Core B can be understood qualitatively using the predictions of hot gaseous phase chemical models \citep[e.g.,][]{2003A&A...412..133V}.
Panel (d) of Figure \ref{fig:f1} shows a ring-like distribution of $^{33}$SO around Core A.
Such a structure has been found for the first time in this source owing to the high spatial resolution.
In this subsection, we discuss such this $^{33}$SO structure.

\begin{figure}[!th]
\figurenum{11}
 \begin{center}
  \includegraphics[bb = 0 10 206 195, scale=1.0]{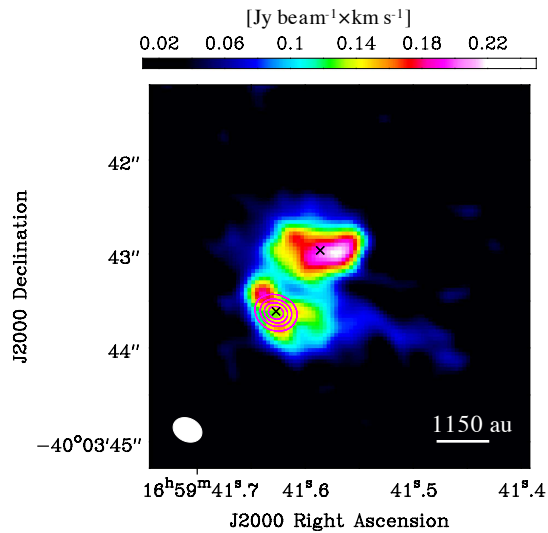} 
 \end{center}
\caption{Comparison of spatial distributions of $^{33}$SO (color) and H30$\alpha$ (magenta contours; the contour levels are 20, 40, 60, and 80\% of its peak intensity of 11.1 Jy beam$^{-1}$$\times$ km s$^{-1}$). Black crosses indicate positions of Core A and Core B. \label{fig:f5}}
\end{figure}

Figure \ref{fig:f5} shows the spatial distributions of $^{33}$SO (color) and H30$\alpha$ (magenta contours) emissions.
The strong emission peaks of $^{33}$SO are located at the outer edge of H30$\alpha$ emission.
There are two possible scenarios for such a ring-like structure around the HC\ion{H}{2} region; one is molecular destruction by the UV radiation from the central star, and the other is the gas-phase formation of SO at this position.
If such a ring-like structure has been formed by molecular destruction by the UV radiation from the central star, other molecules are expected to show similar structures.
However, we do not find ring-like structures in moment zero images of other molecules (Figures \ref{fig:f2} and \ref{fig:f3}).
This implies that the destruction did not produce the ring-like structure of $^{33}$SO emission by the UV radiation from the central star.

Sulfur monoxide (SO) is considered as a shock tracer. 
At Core A, shocks can be produced by a molecular outflow and an expanding motion of the HC\ion{H}{2} region.
However, the morphology of the $^{33}$SO emission does not match the orientation of the molecular outflow nor jet reported by \citet{2011ApJ...736..150G}.
Hence, the molecular outflow does not seem to be an origin of the $^{33}$SO emission feature.
In summary, $^{33}$SO is expected to be enhanced in shock regions produced by an interaction between a thick cloud and an expanding motion of the HC\ion{H}{2} region.
In shock regions, SO is considered to be formed by the following reactions \citep{2014AA...567A..95E}:
\begin{equation} \label{equ:S1}
{\rm O}_{2} + {\rm S} \rightarrow {\rm {SO}} + {\rm O},
\end{equation}
and
\begin{equation} \label{equ:S2}
{\rm {OH}} + {\rm S} \rightarrow {\rm {SO}} + {\rm H}.
\end{equation}
\citet{2014AA...567A..95E} suggested that the reaction (\ref{equ:S1}) is efficient just after the shock passes and the temperature is still high ($T \geq 1000$ K), while the reaction (\ref{equ:S2}) becomes more efficient in cool gas regions.
According to the Kinetic Database for Astrochemistry (KIDA: \url{http://kida.astrophy.u-bordeaux.fr/}), the $\alpha$ values for the reaction rate coefficients, defined as $k = \alpha(T/300)^{\beta} {\rm {exp}}(-\gamma/T)$ cm$^{3}$ s$^{-1}$, are $2.1 \times 10^{-12}$ and $6.6 \times 10^{-11}$ for reactions (\ref{equ:S1}) and (\ref{equ:S2}), respectively.
The $\gamma$ values for both of the reactions are reported as zero \citep{2017MNRAS.469..435V}.
Therefore, the reaction (\ref{equ:S2}) is expected to proceed faster than the reaction (\ref{equ:S1}), and could be a main formation pathway of SO around the observed HC\ion{H}{2} region.

\section{Conclusion} \label{sec:con}

We have analyzed the ALMA cycle 2 data toward the IRAS 16562--3959 high-mass star-forming region.
We spatially resolve the central bright sources into a binary system, a HC\ion{H}{2} region and a younger molecule-rich core.
We identified molecular emission cores using the moment zero images of the H30$\alpha$ line and a CH$_{3}$OH line, which we name Cores A, B, and C.
We have detected several oxygen-bearing COMs, CH$_{3}$CN, and HC$_{3}$N and derived their column densities and excitation temperatures at three cores.
While oxygen-bearing COMs have been detected toward Cores B and C, CH$_{3}$CN, HC$_{3}$N, and HNCO are located in all of the cores.

We compare the chemical composition between Core B and Core C.
The fractional abundances at Core B are higher than those at Core C by around one order of magnitude, while the fractional abundances of CH$_{3}$OH and H$_{2}$CO at Core B are higher by a factor of $\simeq 100$.
These results seem to imply that Core B is a more evolved hot core, where enough time has elapsed for molecules to sublimate from dust grains and new molecules to form in the hot gas.

We investigate the main formation mechanism of H$_{2}$CO toward this high-mass star-forming region by a comparison of the spatial distributions between H$_{2}$CO and C$^{18}$O.
Their extended arc-like structure suggests that gas-phase reaction and/or a grain surface reaction followed by non-thermal evaporation are likely formation routes of the gas-phase H$_{2}$CO.
The enhancement of the gas-phase H$_{2}$CO at Core B also indicates efficient thermal evaporation of H$_{2}$CO.

The spatial distribution of $^{33}$SO around Core A shows a unique structure distributing at the outer edge of the H30$\alpha$ emission region.
These results seem to indicate that $^{33}$SO is enhanced in a shock region produced by an expanding motion of the HC\ion{H}{2} region.

\acknowledgments
This paper makes use of the following ALMA data: ADS/JAO.ALMA\#2013.1.00489.S. 
ALMA is a partnership of ESO (representing its member states), NSF (USA) and NINS (Japan), together with NRC (Canada), MOST and ASIAA (Taiwan), and KASI (Republic of Korea), in cooperation with the Republic of Chile. 
The Joint ALMA Observatory is operated by ESO, AUI/NRAO and NAOJ.
Based on analysis carried out with the CASSIS software and JPL and CDMS spectroscopic databases. 
CASSIS has been developed by IRAP-UPS/CNRS (\url{http://cassis.irap.omp.eu}). 
This work was supported by JSPS KAKENHI Grant Number JP20K14523.

%

\vspace{5mm}
\facilities{Atacama Large Millimeter/submillimeter Array (ALMA)}


\software{Common Astronomy Software Applications package \citep[CASA;][]{2007ASPC..376..127M}, CASSIS \citep{2011IAUS..280P.120C}}



\appendix

\section{H30$\alpha$ spectra toward Core A} \label{sec:Halpha}

\begin{figure}[!th]
\figurenum{12}
 \begin{center}
  \includegraphics[bb = 0 30 324 235, scale=0.75]{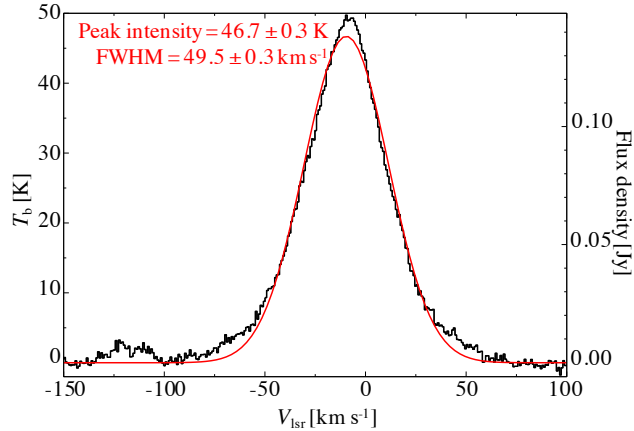} 
 \end{center}
\caption{Spectra of H30$\alpha$ toward Core A. Red line indicates the result of a Gaussian fitting. \label{fig:H}}
\end{figure}

Figure \ref{fig:H} shows a spectra of the H30$\alpha$ line (231.9009 GHz) toward Core A.
A Gaussian fitting to the line is shown with a red line in Figure \ref{fig:H}.
The peak intensity and full width half maximum (FWHM) were derived to be $46.7 \pm 0.3$ K ($1\sigma$) and $49.5 \pm 0.3$ km s$^{-1}$, respectively.

Under LTE  conditions, 
the line optical depth  of the recombination line is given by 
$
\mathcal{T}_{L}\phi(\nu),
$
where $\mathcal{T}_L$ is defined in Equation (B5) in \citet{2014ApJ...796..117G}  and 
$\phi(\nu)$ is the line profile. 
Dividing $\mathcal{T}_L$ by the free-free opacity $\tau_{ff}$ \citep[e.g.][Section 10.6]{2013tra..book.....W} we obtain the
line to continuum equivalent width. For the H30$\alpha$ transition and  assuming $T_e=7000$ K, this width is 104.044 MHz or 134.504 km s$^{-1}$.
 
Therefore, assuming optically thin conditions and considering that the line is well fitted by a Gaussian (which means we can ignore to a first approximation pressure and opacity broadening), the line peak to continuum ratio for a $49.5 \pm 0.3$ km s$^{-1}$ width H30$\alpha$ line is \[\frac{134.504\, \text{km s$^{-1}$}}{\sqrt{\pi/\log(16)}\, 49.5\,\text{km s$^{-1}$}} = 2.55~~.\] Therefore, the expected free-free contribution to the continuum is $46.7~{\rm K}/2.55 = 18.3$ K.

\section{Model results of H$_{2}$CO} \label{sec:H2COmodel}

We investigate main formation pathways of H$_{2}$CO using the results of \citet{2019ApJ...881...57T} with the Nautilus \citep{2016MNRAS.459.3756R}.
The details of this model were described in \citet{2019ApJ...881...57T}.
The gas density ($n_{\rm H}$) increases from $10^{4}$ to $10^{7}$ cm$^{-3}$.
In order to investigate the temperature dependence in detail, we use results of the 3-phase (gas, dust surface, and bulk of ice mantle) and slow warm-up ($1 \times 10^{6}$ yr) model.  

\begin{figure}[!th]
\figurenum{13}
\begin{center}
\includegraphics[bb = 0 250 595 800, scale = 0.68]{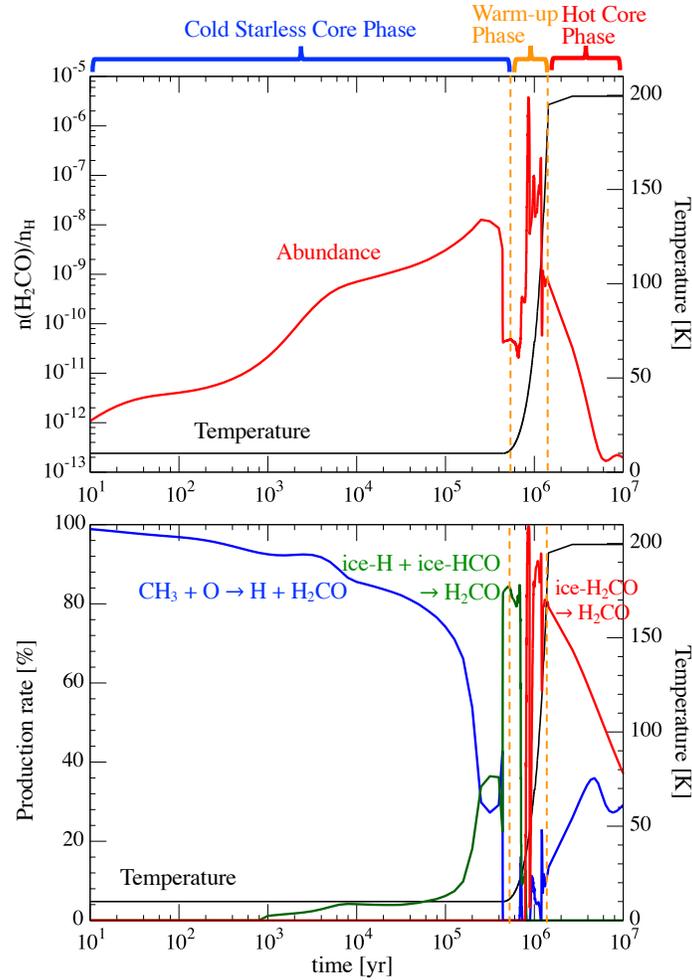} 
\end{center}
\caption{Time dependences of the gas-phase H$_{2}$CO abundance (upper panel) and fractions of its main formation pathway (lower panel). Black lines indicate the time dependence of the temperature. \label{fig:H2COmodel}}
\end{figure}

Figure \ref{fig:H2COmodel} represents the time evolution of H$_{2}$CO from a starless core to the hot core phase through the warm-up phase.
The upper panel of Figure \ref{fig:H2COmodel} shows the time evolution of gas-phase H$_{2}$CO abundance together with the temperature profile.
The lower panel shows the production rates of the main formation pathways of H$_{2}$CO together with the temperature profile.

During the low-temperature (10 K) starless core stage, the gas-phase reaction between CH$_{3}$ and oxygen atom (O) is the main formation route of H$_{2}$CO.
After the gas density reaches $10^{7}$ cm$^{-3}$ and the temperature starts to increase, the following reaction contributes to the gas-phase H$_{2}$CO formation:
\begin{equation} \label{equ:H2COdust1}
{\rm {ice-H}} + {\rm {ice-HCO}} \rightarrow {\rm {H}}_{2}{\rm {CO}}.
\end{equation}
This reaction efficiently works before the temperature reaches at 25 K.
After the temperature rises above $\sim 50$ K, the thermal evaporation of H$_{2}$CO, which is formed by successive hydrogenation reactions of CO molecules on dust surfaces, is the most efficient route.

\end{document}